\documentstyle[twoside,fleqn,espcrc2,epsf]{article}
\newcommand{\AmS}{{\protect\the\textfont2
  A\kern-.1667em\lower.5ex\hbox{M}\kern-.125emS}}
\def\cancel#1#2{\ooalign{$\hfil#1\mkern1mu/\hfil$\crcr$#1#2$}}
\def\slash#1{\mathpalette\cancel{#1}}
\title{Anisotropic Lattices and Dynamical Fermions
\thanks{This work was conducted on the QCDSP machines at 
Columbia University and RIKEN-BNL Research Center. 
TM and LL are supported by DOE. 
}}

\author{L. Levkova\address{Department of Physics, Columbia 
University, New York, NY, 10027} and
T. Manke$^{\rm a}$}
\begin{document}
\bibliographystyle{apsrev}

\begin{abstract}

We report results from full QCD calculations with two flavors of
dynamical staggered fermions on anisotropic lattices.  The physical
anisotropy as determined from spatial and temporal masses, their
corresponding dispersion relations, and spatial and temporal Wilson
loops is studied as a function of the bare gauge anisotropy and the
bare velocity of light appearing in the Dirac operator.  The
anisotropy dependence of staggered fermion flavor symmetry breaking is
also examined.  These results will then be applied to the study of
2-flavor QCD thermodynamics.

\vspace{1pc}
\end{abstract}

\maketitle

\section{INTRODUCTION}
Anisotropic lattices have been used extensively for $T=0$
calculations\cite{zeroT}. Finite temperature calculations using anisotropic lattices
exploit the natural asymmetry of finite temperature field theory to reduce lattice spacing errors
associated with the transfer matrix at less cost than is required for the full
continuum limit\cite{finT}.

With a sufficiently small value for the temporal lattice spacing, 
$a_t$, we can vary the temperature in small discrete steps by varying the number of time slices,
$N_t$.  In this study  we sweep 
through the transition by studying different $N_t$'s (8--64). %, 12, 16, 20, 24 and 64.
 Varying the temperature at fixed temporal and spatial lattice
spacing separates temperature and lattice spacing effects, allowing
a study of the temperature dependence with all other parameters fixed.%fixed lattice spacing errors.

The coefficients needed to relate lattice observables to the physical energy and pressure
are determined as a by-product of the zero temperature studies needed
to choose the bare parameters. %$\beta$, the bare gauge and fermion anisotropies and the quark
%mass.  
Once determined, these ``Karsch'' coefficients\cite{karsch} can be used for all 
temperatures since they depend only on the intrinsic lattice
parameters and not on $N_t$.  This allows a straight-forward determination 
of the temperature dependence of the energy and pressure, again at fixed
lattice spacing.
With two or more slightly different values for $a_t$, a high-resolution 
sampling of temperatures can be investigated. 

As the temporal lattice spacing, $a_t$, approaches 
the continuum limit $(a_t\rightarrow 0)$, the part of the flavor symmetry, which is violated by terms  
 of ${\cal O}(a_t^2)$, is expected to be restored. 
In this study we are examining our data for evidence of improvement of the 
flavor symmetry, when $a_t$ becomes sufficiently small.
%From our data we see 
%evidence for improvement of the flavor symmetry, when $a_t$ becomes sufficiently small.

%While this approach naturally reduces finite lattice spacing errors
%associated with $a_t$, we plan to include improvements to the spatial
%parts of the staggered fermion action in future work so that the
%${\cal O}(a_s^2)$ errors are reduced as well.

\section{THE ANISOTROPIC STAGGERED ACTION}
We are simulating full QCD with two dynamical flavors of staggered fermions on an anisotropic lattice.
Our calculations are based on the QCD action
$S^{\xi} = S_G^{\xi} +  S_F^{\xi}$,  where the gauge action is:
\begin{eqnarray}
\label{eq:asym_gauge_action}
S_G^{\xi} 
  = -\frac{\beta}{N_c}\frac{1}{\xi_o}\left[
     \sum_{x, s>s^\prime} P_{ss^\prime}(x)  +
  \xi_o^2 \sum_{x, s} P_{st}(x) \right],
     \nonumber
\end{eqnarray}
and the fermion action is:
\begin{eqnarray} 
\label{eq:aniso_lattice_quark_action}
S_F^{\xi}
  &=& \sum_{x} \overline{\psi}(x) \left[
      m_f + 
      \nu_t  \slash{D}^{\rm Staggered}_t\right] \psi(x)\nonumber\\
& & +
     \sum_{x} \overline{\psi}(x)\left[  \frac{1}{\xi_0} \sum_s \slash{D}^{\rm Staggered}_s 
      \right]  \psi(x) . \nonumber
\end{eqnarray}

In our simulations we attempt to examine the
QCD phase transition for volumes ($16^3\times4$), quark masses ($m_\pi/m_\rho \approx 0.3$) and a spatial lattice
spacing ($a_s \approx 0.3$ fm) similar to those used in $N_t=4$, 2-flavor thermodynamic studies on isotropic lattices.
Thus, we adjust the bare bare anisotropy ($\xi_0$) and the renormalization of the speed of light ($\nu_t$) to yield the required 
$a_s$ and $m_\pi/m_\rho$ but work with a much smaller temporal
lattice spacing choosen so the critical value of $N_t$ is approximately 16. 

\section{SIMULATIONS}
%\vspace{-1cm}
\begin{table}[h]
\begin{small}
\begin{tabular}{llrlll}  \hline 

{\em run}& {\em volume} & {\em traj.}& $\beta$ & $\xi_0$ & $m_f$ \\ \hline

1 & $16^3$x32 &  5800   & 5.425   & 1.5    &      0.025 \\ 

2 & $16^2$x24x32 & 5100 &  5.425 &     1.5      &  0.025\\

3 & $16^2$x24x64 & 1300 &  5.695  &  2.5   &   0.025\\ 

4 & $16^2$x24x64& 1400 &  5.725 &   3.44  &   0.025 \\ 

5 & $16^2$x24x64 & 3400 & 5.6 &   3.75   & 0.025 \\ 

6 & $16^2$x24x64 & 3200  & 5.3 & 3.0 & 0.008 \\  

7 & $16^3$x24 & 3500  & 5.3 & 3.0 &  0.008 \\ 

8 & $16^3$x20 & 2800  & 5.3 & 3.0 & 0.008 \\  

9 & $16^3$x16 & 4500     &5.3 & 3.0 &0.008\\

10 & $16^3$x12 & 8200  & 5.3 & 3.0 & 0.008 \\

11 & $16^3$x8 & 5500  & 5.3 & 3.0 &  0.008 \\ \hline 

\end{tabular}
\end{small}
%\vspace{0.1cm}
\caption{Parameters of all the calculations. All runs have dynamical $\nu_t=1.0$ except run 3 which has $\nu_t=1.2$.}
%\vspace{-0.6cm}
\end{table}
%\vspace{-0.1cm}

We have used the zero temperature runs 1-6 for scale-setting and determination of the anisotropy. 
%to determine masses and to measure anisotropies from both hadron masses and Wilson loops.
For all the zero-temperature runs the renormalized anisotropy $\xi_r$
is calculated both from the $\rho$ masses in the spatial and temporal directions and from matching the static potentials\cite{match}. 
Figure 1 shows that both methods provide values for $\xi_r$ which are reasonably close.

The idea behind run 7 through 11 is that we keep the spatial lattice spacing, $a_s$, and all other run parameters
constant and change only the number of lattice points, $N_t$, in the temporal direction.
%This way the physical volume is kept constant while the temperature is changed and  we gradually sweep through 
%the phase transition. The results from $<\bar{\psi}\psi>$ measurements are presented in figure 4.

Our simulations implement the R-algorithm\cite{R-alg} with step--size $\Delta t =0.005$ for all jobs from Table 1.
%We have performed a step--size test whether this choice enssures that our simulations 
%are done in the stable regime of the R-algorithm where the finite  step--size errors are smaller than
%the errors on the physical quantities.
%On volume $8^3\times 32$ and parameters close to those of run 7 (where we simulate thermodynamics)
%we measure $<\bar{\psi}\psi>$ and $\frac{m_\pi}{m_\rho}$ for three $\Delta t$'s. 
Figure 2 shows that our choice of $\Delta t$ is consistent with the requirements for small finite step--size
 error on the physical quantities.

\section{THE VELOCITY OF LIGHT}
The velocity of light on a lattice can be defined through the meson dispersion relation:
\begin{eqnarray}
E_{t, phys}^{2}(p_s)&=& \frac{E_{t, lat}^{2}(0)}{a_t^2}+ c_{ts}^2(p_s)P^{2}_{s, lat}\frac{1}{a_s^2} \nonumber
%E_s^{2, phys}(p_{s^\prime})&=& \frac{E_s^{2, lat}(0)}{a_s^2}+ {c_{s{s^{\prime}}}}^2(p)P^{2, lat}_{s^\prime}\frac{1}{a_s^2} \nonumber\\
%E_s^{2, phys}(p_t)&=& \frac{E_s^{2, lat}(0)}{a_s^2}+ c_{st}^2(p)P^{2, lat}_t\frac{1}{a_t^2} \nonumber
\end{eqnarray}

We tune $\nu_t$, so that $c_{ts}(p_s) \approx 1$. %This should automatically 
%tune the velocity of light in the rest of the dispersion relations to be $\sim 1$ if there is 
%consistency with the effective theory in the continuum limit (some tests of this were done).
The velocity of light is calculated for the $\pi$ propagating in the temporal direction 
and having non--zero momentum for three values of $\nu_t$: 1.0, 0.8 and 1.2,
 where the first is a dynamical parameter and the last two are valence parameters.
From Figure 3 we see that the choice of $\nu_t= 1.0$ gives velocity of light closest to 1.0. 

\section{IMPROVEMENT OF THE FLAVOR SYMMETRY}

Table 2 shows the meson masses for run 3 ($a_s=0.231(7)$ fm, $a_t=0.072(4)$ fm) 
and run 4 ($a_s=0.243(3)$ fm, $a_t = 0.057(2)$ fm). 

We choose $\Delta_\pi=(m_{\pi_2} - m_\pi)/m_\rho$, where $\pi_2$ is the second local staggered pion, as a quantitative measure of the flavor symmetry breaking in the spatial (S) and temporal (T) directions. The data shows that in the temporal direction for both runs $\Delta_\pi$ is smaller than its value in the spatial direction, which means that we are seeing improvement of the flavor symmetry as $a_t$ becomes 
finer.
Especially for run 4, the $\pi$ and $\pi_2$ look virtually
degenerate.

\begin{table}[h]
\begin{small}
\begin{tabular}{@{}lllll}   \hline 

{\em am}          &        {\em S, \#3}     &   {\em T, \#3} & {\em S, \#4} & {\em T, \#4}\\ \hline

 $am_\pi$                &  0.680(3)          &    0.214(2) &     0.771(4) &     0.181(2)    \\

 $am_{\pi_2}$             &   0.759(12)         &  0.219(2) &       0.836(21) & 0.181(2)  \\

 $am_\rho$              &   0.902(26)          & 0.283(4) &   0.948(13) &  0.222(4) \\ 

$\Delta_\pi$       &0.09(1)   & 0.017(9)   &  0.07(2)  &  0.000(2)   \\ \hline 
\end{tabular}
\end{small}

\caption{Meson masses for run 3 and 4.}
\end{table}

\section{THE PHASE TRANSITION}
%After the zero-temperature scale--setting calculations, we have chosen the most apropriate parameters for our 
%purpose to simulate thermodynamics: $\beta = 5.3$, $\nu_t=1.0$, $\xi_0=3.0$, $m_f=0.008$. 
The thermodynamics runs 7--11 have volume $16^3\times N_t$, where $N_t=$ 8, 12, 16, 20 and 24. 
The bare parameters are kept constant for all runs 6--11. Figure 4 shows the sweep through the phase transition as we gradually change the temperature by varying only $N_t$. %This appears less sharp than corresponding isotrpoic resultas[].
%We can estimate from the inflection point of the fitting {\bf tanh}: $T_c \approx 160$ MeV.
We fit our data to a hyperbolic tangent form and determine $T_c \approx 160$ MeV from the inflection point.

\section{CONCLUSIONS}
We have studied the finite temperature QCD phase transition 
using staggered fermions on an anisotropic lattice with 
anisotropy of $\approx 4$.  This allows us to explore the 
temperature dependence of the transition with all other parameters 
fixed.  The results are roughly consistent with earlier, 
isotropic studies, showing a value of the critical temperature 
of approximately 160 MeV.
While this approach naturally reduces finite lattice spacing errors
associated with $a_t$, we plan to include improvements to the spatial
parts of the staggered fermion action so that the
${\cal O}(a_s^2)$ errors are reduced as well.

%\item{$T_c \approx 160 MeV$}
%\item{Improvement of the flavor symmetry as the temporal lattice spacing, $a_t$, aproaches
%the continuum limit $(a_t\rightarrow 0)$.}
%\item{With two or more slightly different values for $a_t$, a high-resolution sampling
% of temperatures can be investigated.}
%\item{Spatial improvements are also possible (not implemented in our simulations yet)}
%\item{Studying the temperature dependence of the EOS by varying only $N_t$ and keeping the lattice
% spacing and all other parameters fixed.}
%\item{Difficulty: Many parameters to tune. But once tuned at zero-temperature we can sweep
% through the phase transition without changing any of them except $N_t$.}

%\end{itemize}

\begin{figure}[th]
\epsfxsize=\hsize
\epsfbox{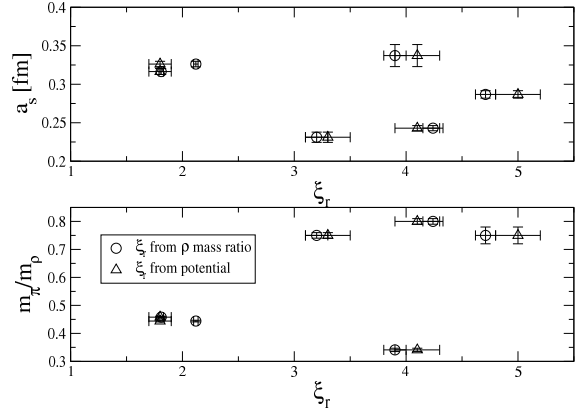}

\caption{Scatter plots of the zero temperature runs 1--6.}

\end{figure}

\begin{figure}[bh]
\epsfxsize=\hsize
\begin{center}
\epsfbox{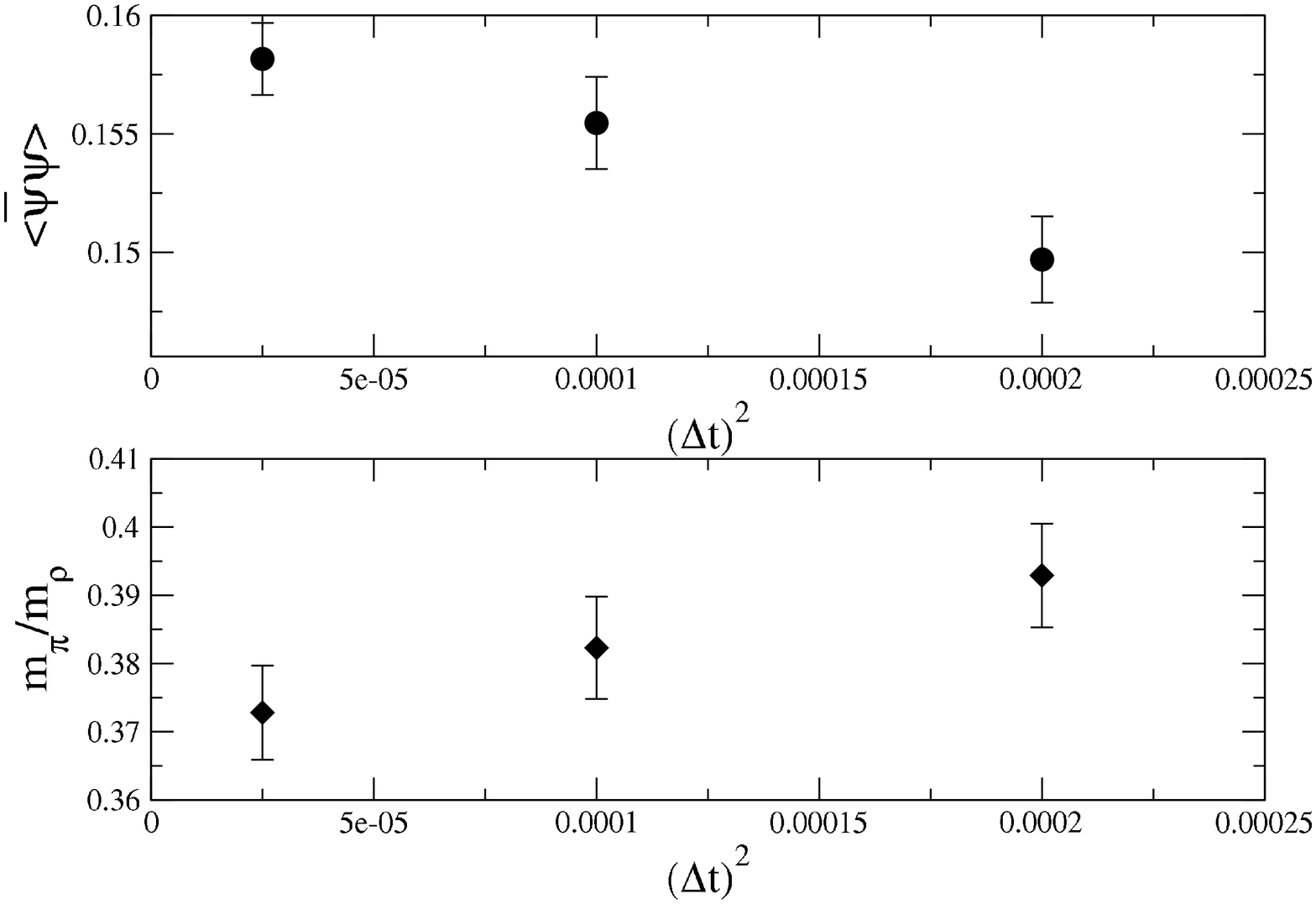}
\end{center}
\caption{Step--size test. Volume $8^3\times 32$, $\beta = 5.35$, $\nu_t=1.0$, $\xi_0=3.5$, $m_f=0.006$.} 
\end{figure}

\begin{figure}[th]
\epsfxsize=\hsize
\epsfbox{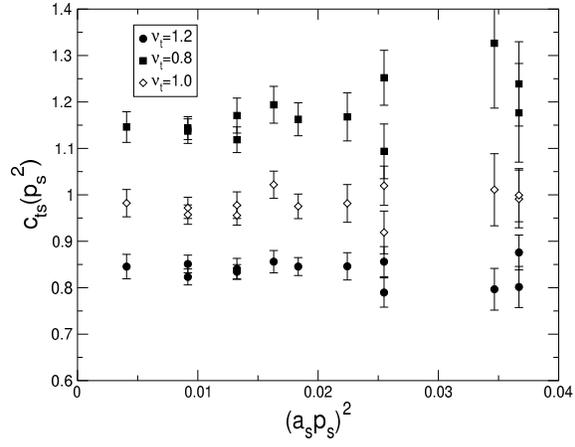}
\caption{Tuning of the velocity of light using a  $16^2\times24\times 64$ volume, $\beta = 5.3$, $\xi_0=3.0$, $m_f=0.008$.}
\end{figure}
\vspace{-6cm}
\begin{figure}[thb]
\begin{center}
\epsfxsize=\hsize
\epsfbox{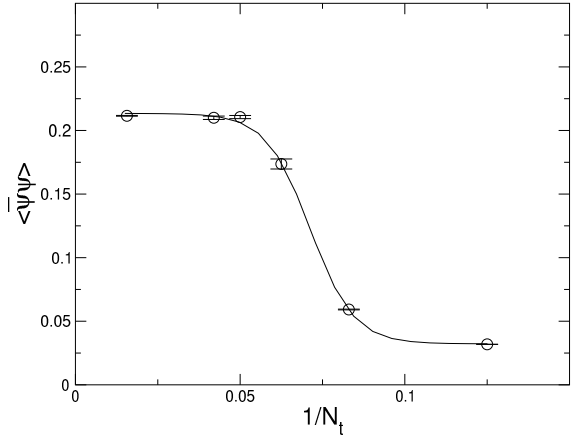}
\end{center}
\caption{The temperature dependence of $\langle \overline{\psi}\psi\rangle$ in the region of $T_c$.}
\end{figure}

\end{document}